       \documentstyle [12pt]   {article}
 \evensidemargin\oddsidemargin
\begin{document}
\count0 = 1
%\begin{titlepage}
 \title{{ Restrictions on Information Transfer\\ 
 in Quantum Measurements     \\ }}
\small\author{S.N.Mayburov \\
Lebedev Inst. of Physics\\
Leninsky Prospect 53, Moscow, Russia, 117924\\
E-mail :\quad   mayburov@sci.lebedev.ru}
\date {}
\maketitle
\begin{abstract}

  Information-theoretical restrictions on the
 information, transferred in quantum measurements,
are  regarded for
%       In particular G$\it\ddot{o}$del information incompleteness
 the measurement of quantum object $S$ performed via its interaction with 
   information system $O$. 
This information restrictions,  induced by Heisenberg commutation relations,
  are  derived in the formalism of inference maps in
Hilbert space. 
 $O$ restricted states $\xi^O$ are calculated from
 Shr$\ddot {o}$dinger $S$,$O$ dynamics
 and the structure of $O$ observables set (algebra);
 $O$ decoherence by its
environment is  also accounted for some $S$, $O$ systems.
%           $R_O$  'preferred' basis $|O_j\rangle$ defined by
%           $O$ state decoherence via $O$ - E interactions.
% The analogous  restrictions derived in Algebraic QM  formalism
% from the consideration of Segal algebra $\cal U_O$  of $O$ observables;
%the resulting set of  $O$  restricted states $\{\xi^O_i\}$
% is the duallinear  space of  $ \cal U_O$. From
% Segal theorem for associative  (sub)algebras.
  It's shown that this principal constraints on the information transfer
  result in
the stochasticity of measurement outcomes;  consequently,
   $\xi^O$ describes the  random 'pointer'
outcomes $q_j$  observed by $O$ in the individual events.
\end{abstract}
\vspace{10mm}
% $ \quad \quad $ {\bf Key Words: \,}
% \keywords {Quantum Measurements; Information Transfer}
\small  
% \quad   Int. J. Quantum Information 5, 279 (2007) \\}
% \small {  \qquad  Torino,  May 2006 }
%
\quad \quad {\it Presented at Symposium on Foundations of Modern Physics}
\\
\vspace {2mm}
 \quad \quad {\it \quad \quad Vienna, june 2007} 
\vspace{16mm}
%\end{titlepage}

%\begin {sloppypar}
%
%
%

\section  {Introduction}

Despite its significant achievements, Measurement Theory of quantum mechanics
 (QM) still contains some open questions concerned with its internal
 consistency$^{1,2,3}$. The most famous and oldest  of them is
% There are still  several unsettled problems
% concerning the interpretation of
% Quantum Mechanics (QM), and the majority of them
% involves to some extent the quantum measurement processes
% (Jauch,1968; Aharonov,1981).
% is a universally acknowledged
% theory, yet there are still several unsolved questions
% concerning its interpretation.
% The oldest and most prominent of them is probably
  the State Collapse or Objectification
   problem, but there are also others, more subtle and
 less well-known$^{4,5,6}$.
%  (Jauch,1968;  D'Espagnat,1990).
% It was proposed that this problems should be analyzed not only
% within Quantum Mechanics (QM) domain only,
%  but in more  wide context involving
% Information Theory, Logic and Methamathematics$^{5,6}$.
%    (Breuer,1996; Mittelstaedt, 1998).
In this paper this problem called often the Quantum Measurement problem
  is studied mainly 
within the framework of Information Theory.
% with complete account of Quantum
% Dynamics results for Measuring System (MS).
 % which in principle can lead to   the important conclusions.
Really,   the  measurement of some system $S$ 
includes   the  transfer of information from
 $S$    to the information system  $O$  (Observer) which processes
 and memorizes it. In Information Theory, any measuring system (MS)
can be regarded as the information transferring channel, which transfers
the parameters of  $S$ state
to $O$.
% 
% via this channel$^7$.
%  via  S interaction with measuring system (MS) which element is $O$
% (Guilini,1996; Duvenhage,2002).
Consequently, the possible   restrictions on the information
 transfer   from $S$ to $O$  can influence
  the effects  observed in the  measurements.
   In our previous paper it was shown that such
 restrictions are principally important for the whole picture 
of measurement. In particular, they
 induce the unavoidable stochasticity in the observed by $O$
outcomes of measurements, which, by the all appearances, coincide
with the collapse of measured state$^8$.
  Our calculations of this effects 
 exploited the formalism of Observable Algebra - $C^*$-algebra which is  most
general and deep mathematical QM formulation$^4$.
 However, this formalism is rather complicated and abstract, 
 so in this paper the same calculations
 are reconsidered by means of standard   Schr$\ddot o$dinger
 QM formalism;
% not including Reduction postulate which is the aim of
%   features of information transfer in more detail and
it permits also to analyze
the fundamental QM aspects more simply and straightforwardly. 
 It supposed that  the evolution of
all objects, including macroscopic ones, like $O$, can be described
 by the quantum state (density matrix) $\rho(t)$ which  obeys to
 Schr$\ddot o$dinger-Lioville equation in arbitrary reference frame (RF). 
The information-theoretical approach
 of system self-description or 'measurement from inside' is applied
 to the consistent description of information acquisition in
quantum measurements$^7$.
In  Schr$\ddot{o}$dinger QM framework, the formalism of
 inference maps in Hilbert space$^5$  responds to this approach,
it  applied
  for the calculations of information transfer from $S$ to $O$. 
%  additional analysis of systems observables, in particular, the informational
 
The   principal
 features of  Schr$\ddot{o}$dinger formalism,
 essential for our  measurement formalism, can be formulated as
  QM eigenstates ansatz (): for any
 observable $G$ of quantum finite-dimensional system $\Xi$
there at least two $\Xi$ (pure) states $\rho_i $  for which 
$\Xi$ possess the different real  properties   $g_i$ and they
 are $G$ eigenvalues$^1$. Remind$^4$ that, in general, an arbitrary $\rho$
 will be $G$ eigenstate with eigenvalue $g_a$:
$G,\rho \to g_a$, iff $\bar{G}^l=(\bar{G})^l=g_a^l$ for any natural $l>0$.
Thus,  predicts with definiteness the result of $G$ measurements
for  $\rho_i$ states in the individual events of measurement. All $\Xi$
individual states, i.e. the states in particular event are pure (but can be
unknown), yet the statistical (ensemble) states can be also mixed$^6$.  
As shown below, the set of such $\Xi$ eigenstates $\{ \rho^G_i \}$ for various
$G$ constitutes for  given $\Xi$ the 'information' basis,
 which  permits  to derive the  measurement properties
 of arbitrary state $\rho'$ from its comparison with $\{\rho^G_i \}$
 properties. Overall, 
we shall argue  that  together 
with the systems' self-description formalism of information theory permit
 to construct the consistent measurement formalism, without inclusion of QM
 Reduction Postulate.
Further details concerning with the inference maps,  systems' self-description,
etc., can be found elsewhere$^{6,8}$. Early version of this text
 was published in$^{12}$.

\section                   {Model of Quantum Measurements}

Here our measurement model will be described and some aspects of
QM Measurements Theory, essential for our approach, discussed
 at semiqualitative level.  
% The detailed consideration of Quantum Measurements Theory
%found elsewhere (Busch, 1996), here only its particular aspects
%essential for our  approach are reviewed. To illustrate them,
%the simple  model of measurement will be  used, similar  to von Neuman
%
 In our model MS consists of
  the studied system $S$,  detector $D$ and the information system
   $O$, which memorizes and process the information about MS current state.
The effects of $D$, $O$ decoherence by their environment aren't
of primary importance in our theory, they will be regarded
 briefly in the final part of the paper. 
% n this chapter $O$  isn't regarded as a quantum object in a strict sense
% (it will be done in chap. 3).
% For the simplicity the detector D is dropped in MS chain,
%  D, $O$ decoherence effects will be discussed in the final part of paper.
%    
% Here S,D system is described in fact by Zurek ansatz$^{10}$, quite
% popular for the discussion of QM foundations.
 As in many other  models of measurement$^{1, 10}$,  
   $S$ is taken to be the particle with the spin $\frac {1}{2}$ and
 the measurement of
its projection  $S_z$ is regarded. Its $u,d$ eigenstates
 denoted $|s_{1,2}\rangle$, 
  the  measured $S$ pure state is: 
    $ \psi_s =a_1|s_1\rangle+a_2|s_2\rangle$.
 For the comparison, the incoming
  $u,\, d$ 'test' mixture with the same $\bar{S}_z$ should be regarded 
also. This is $S$  ensemble described by the gemenge$^1$
$W^s=\{|s_i\rangle, \rm{P}_i \}$, where $\rm{P}_i$ $=|a_i|^2$
 are the probabilities
of  individual state $|s_i\rangle$ in this ensemble,
 its statistical state described by the  density matrix:
\begin {equation}
 \rho_s^m= \sum_i |a_i|^2|s_i \rangle \langle s_i|
                                                              \label {AA33}
\end {equation}
% S pure state is represented by
% two-dimensional state vector $\psi_s$.
Analogously to $S$ state, $D$ state in $O$ RF is supposedly  described by
Dirac vector $|D\rangle$ in two-dimensional
% $O$ interaction with environment E - decoherence
% will be  accounted at the next stage of the model regarded in chap. 4.
%     \psi_s=a_1|s_1\rangle+a_2|s_2\rangle
%
%  The pure S,$O$ states  belongs to MS Hilbert space
%  ${\cal H}_{MS}$ defined in  $O'$ RF.
%
%  the density matrixes ${\rho}_{MS}$ set defined on
% $\cal{H}_{MS}$ denoted $\Gamma$.
% The regarded $O$ has one  internal DF
  Hilbert space $\cal {H}_D$. Its basis  constitutes
   $|D_{1,2}\rangle$ eigenstates of $Q$
'pointer' observable with eigenvalues $q_{1,2}$.
%    which for the
%   convenience is taken to be equal to $0,1,2$ correspondingly.
% It assumed also that $|D_{1,2}\rangle$
%constitutes the basis of two dimensional subspace $\cal{H}_D$.
%It's possible for example, if $\cal {H}_O$ states are
%$SU_3$ spinors, then $\cal H'_O$ states will be
%$SU_2$ spinors, so that $|O_{1,2}\rangle$  correspond to
% $u,d$ spin $\frac{1}{2}$ states ( Gasiorowicz,1966).
% $\cal H'_O$ observables conjugated to $Q_O$ are denoted
% $Q_O^x,Q_O^y$.
% $SU_3$ spinors and $|O_0\rangle$ orthogonality to
%$|O_{1,2}\rangle$ in this  model are used mainly for illustrative
%purposes, the same results as below, can be  obtained for $\cal
% H_O=\cal H'_O$ and
%
The initial $D$ state is:            
$|D_0\rangle=\frac{1}{\sqrt{2}}(|D_1\rangle+|D_2\rangle)$.
%
%
% Let us   consider
% the measurement   of   spin projection $\hat{S}_z$,
%  on S   state $\psi_s$;
%  the initial  $O$ state  is  $|O_0\rangle$ and  thus
% |O_o\rangle \, , \label {AAB} \end {equation}
% where $|s_{1,2}\rangle$ are $S_z$ eigenstates
%  with eigenvalues $s_{1,2}=\pm 1$.
% To extract the  studied effect
% it will be compared with result for $|s_i\rangle$ mixture $W^s$
% with the same $s_z$ probabilities $P_i=|a_i|^2$.
%
  $S$, $D$  interaction $\hat{H}_{S,D}$ starts at $t_0$
and finishes effectively at some  $t_1$;
for   Zurek Hamiltonian $H_{S,D}$ with suitable parameters
% Schr$\ddot{o}$dinger equation for $\Psi^{in}_{MS}$
%  the suitable   S$-O$ interaction Hamiltonian$^{10}$ $\hat{H}_I$
  it would result in  $S,\, D$  entangled final  state
$\rho_{S,D}$ or the corresponding  state vector:
\begin {equation}
   \Psi_{S,D} = \sum a_i|s_i\rangle|D_i\rangle 
                                   \label {AA2}
\end {equation}
 relative to $O$ RF. Hence the measurement of
 $S$ eigenstate $|s_{1,2}\rangle$ results in 
$\Psi^{1,2}=|s_{1,2}\rangle |D_{1,2}\rangle$,  this is 
 factorized $S$, $D$ state.
 In  case $a_{1,2}\ne 0$, $D$ also possess the quantum
state $R_D$, but it can't be completely factorized from the entangled
$S$, $D$ state $\Psi_{S,D}$.
% The exact $R_D$ ansatz for individual states will be derived below,
 If to neglect this entanglement, $D$ statistical state coincides with
 MS partial state, obtained by tracing out $S$ degrees of freedom
from $S$, $D$ state. It will be argued that the situation with individual
$D$ state ansatz is more subtle and ambiguous,
 but in our calculations no particular $R_D$ ansatz is used.
%  and thus  $|O_{1,2}\rangle$ are $O$ final states induced by
%   the measurement of eigenstates  $|s_{1,2}\rangle$.
%  for the corresponding .
It turns out that
% \begin {equation}
$ \bar{Q}=|a_1|^2-|a_2|^2$  
% \label {XX2}  \end {equation}
, so $D$ performs  $S_z$ measurement of first kind$^1$.
At $t>t_1$   $D$, $O$ interaction starts and finishes at some $t_2$,
 during this interval,   
some information about $D$ state is transferred to $O$.
 In this chapter $O$ isn't described consistently
 as the quantum object, it will be done in chap. 3. It assumed arbitrarily
 that the acquisition of information by $O$ doesn't violate QM laws.
 It's supposedly true for
 human observer also,  below some terms characteristic for
 human perception will be used in illustrative purposes.

In Information Theory, the signal induced by the measured state and
 registrated by $O$ in event $n$ 
 is characterized by the information pattern (IP) 
$J(n)=\{e_1,... ,e_l\}$, this is the array of numerical parameters, which
represent  the complete signal description available for 
 $O^6$.
 The  difference between two
individual states for $O$ is reflected by the difference of their IPs$^6$;
if some $e_i$ are the same for all studied states, they can be omitted in $J$.
In general, the set of all possible $J$ constitutes the independent
 'information space',
which describes $O$ recognition of  measured states or signals$^7$.
In quantum case, some states parameters can be uncertain,
 below it will be shown that it's unimportant for the corresponding IPs
and only definite $e_i$ should be regarded.
% that they can be substituted by another  definite IP
% parameters called an interference terms. 
 Let's regard first  the measurement of
 $S$ eigenstate $|s_{1,2}\rangle$,  
% $\Psi^{1,2}=|s_{1,2}\rangle |D_{1,2}\rangle$,  this is 
% factorized $S$,D state.
  in that case, 
 $O$ supposedly percepts $|D_{1,2}\rangle$ state in event $n$ as IP:
$$
 J(n)=J^D_{1,2}=e_1=q_{1,2}
$$ 
% - $O$ 'impressions'  (see below).
 For final $|D_{1,2}\rangle$ states their $Q$ eigenvalues $q_{1,2}$, which
 describe $D$ pointer position are $D$ real properties$^1$,
corresponding to the orthogonal projectors $P^D_{1,2}$.
 Hence  for $O$ the  difference between this $D$ states is the objective
or Boolean Difference$^3$ (BD). It means that this states difference
 is equivalent to the distinction between  the logical operands $Yes/No$,
 or that's the same   
between the values $1/0$ of some discrete parameter $L_g$ available
 for $O$. Note  that in QM  all measurable parameters are related to
 the observables which  represented
by Hermitian Operators on $\cal H$ (or POV in general formalism).
 For example,  for  $|D_{1,2}\rangle$ the parameter $L_g=1/0$
is the eigenvalue of projector $P^D_1$.
% expressed by IP $J^D_{1,2}$.
 It will be argued
that such strict correspondence reduces the number of feasible parameters,
 which can be applied
for the discrimination of quantum states.
% From that follows,  as will be shown,  not
%  for all different $D$ states BD would hold$^6$.

     Now let's regard the possible measurement outcomes when
 $a_{1,2} \ne 0$,  i.e. $\psi_s$ is $|s_i\rangle$ superposition.
% consider D (restricted) state $V_d$ related to $\Psi_{S,D}$.
The standard or  'Pedestrian' Interpretation$^2$(PI) of QM  claims that
  without the inclusion of Reduction Postulate into QM formalism, 
 $O$ should percept  $S$, $D$ entangled state $\Psi_{S,D}$ as 
  the  superposition of IPs induced by $|D_{1,2}\rangle$,
 i.e. the simultaneous coexistence of two IP $J^D_1$ and
 $J^D_2$. This hypothesis is the essence of famous
 'Schr$\ddot{o}$dinger Cat' Paradox$^3$.
 More realistically, one can expect   at least  that  
  IP $J^s$, induced by $\Psi_{S,D}$, would differ for $O$
 from   $J^D_{1,2}$.
% But  because it isn't observed experimentally, PI supporters conclude that
% Reduction Postulate
% should be added to QM formalism$^9$.
 Yet the situation isn't so simple
 and doesn't favor such prompt jump  to the  conclusions. 
% assumes that in any event $n$
% $O$ can discriminate  $V_d$ from $|D_1\rangle$ and $|D_2\rangle$,
% but  each $|D_i\rangle$ is  unambiguously 
% characterized by the parameter $q_i$. 
Really, given PI implications are correct, $O$ should distinguish
 in a single event
$\Psi_{S,D}$, i.e. the corresponding $D$ state  $R_D$
 from each $|D_{1,2}\rangle$. 
Hence  the relation of corresponding $O$ IPs should be characterized
 by BD, i. e. for $\Psi_{S,D}$  the corresponding $J^s \ne J^D_{1,2}$.
  Hence 
  it should  be  at least  one  measurable $D$ parameter $G^D$
  which value  $g_0$ for $\Psi_{S,D}$
is different from its values $g_{1,2}$ for $|D_{1,2}\rangle$.
% which values
%  for $V_d$ and $|D_i\rangle$ are different - i.e. to obey
% Boolean distinguishability (BD) (Jauch).   
% However in QM all measurable parameters are related to observables - 
% Hermitian Operators,
To verify this hypothesis for $\Psi_{S,D}$ and $|D_{1,2}\rangle$,
 one should check the set (algebra) of
   $D$ PV observables $\{ G^D \}$ as possible candidates
 (the role of joint $S, \,D$ observables regarded below).
% since in QM all feasible parameters
% correspond to QM observables.
% The possibility that such observable is $Q$, 
%  $P^D_i$ or some their linear form
% is excluded beforehand,  because
% $Q$ has only two eigenvalues $q_{1,2}$  occupied by $|D_{1,2}\rangle$.
The simple check shows that no  $D$ observable also can satisfy
 to this demands; POV generalization of this ansatz will be discussed below,
but it doesn't change this conclusion. 
% difference, the reason of it is that $P^D_i$ is the only
% $|D_i\rangle$ projector
% and $R_D$ doesn't have the projector on $\cal H_D$
Really, suppose that such $G^D$ - Hermitian operator exists,
% and its eigenvalues $g_0\ne g_{1,2}$.
then it follows:
\begin {equation} 
   G^D\Psi_{S,D}=a_1|s_1\rangle G^D|D_1\rangle+a_2|s_2\rangle G^D |D_2\rangle
    =g_0\Psi_{S,D}
                      \label {E1}
\end {equation}
As easy to see, for $D$ observables such equality fulfilled only for $G^D=I$. 
% eigenvalue is the same for all $D$ states.
It can be shown that any nontrivial $G^D$ with such properties  
can respond only to the  nonlinear operator on $\cal H_D$,
 hence the observation of such difference is incompatible
 with standard QM formalism. 
Consequently,  it's impossible for $O$ to
distinguish $|D_i\rangle$ from $R_D$ i.e. from $\Psi_{S,D}$ of (\ref {AA2})
 in a single event. Meanwhile,
for  $S$ ensemble one can
expect that the correct $\bar Q$ is obtained by $O$ in $S$ detection, 
to fulfill this condition, $O$ should observe 
the stochastic $q_{1,2}$ outcomes with probabilities $\rm{P}_{1,2}$.
 This considerations put doubts on the necessity of
 independent Reduction Postulate
in QM, the similar hypothesis was proposed first by Wigner$^9$ 
from the considerations of quantum measurements and consequent
information aquisition by human observer.
% One can regard this result as the reason
% to introduce Reduction Postulate in QM, however,
To give more arguments in favor of our theory, it's
  instructive to consider in detail  the  possible influence 
 of $O$ quantum properties on the measurement picture.
Note that the obtained results don't mean that $R_D$ is the probabilistic
 mixture of $|D_{1,2}\rangle$, rather $R_D$ can be characterized
 as their 'weak' superposition, induced by the entanglement of
 $S$, $D$ states.   

% this semiqualitative  analysis prompts to perform more detailed account
% of $O$ quantum properties, before any conclusions can be done.
% which is performed in the next chapter.
% Besides it illustrates
% the intricated relations between Dynamics and Information
% which arise in quantum measurements.  
%               because $|D_i\rangle$ has the only. 

\section {Quantum Measurements and System Self-description } 

Now the information system  $O$ will be regarded as the quantum object also,
so MS is described by a quantum state $\rho_{MS}$ relative to  some other
 RF $O'$.
% This formalism permits to connect the information-theoretical
% $O$ IP ansatz with the quantum dynamics of MS system.
 We shall regard 
 $O$ with the same internal structure as  $D$ posess: $O$ pure state is
 a vector in two-dimensional Hilbert space $\cal H_O$. Analogously to $D$,
we settle $O$ initial 
state $|O_0\rangle=\frac{|O_1\rangle+|O_2\rangle}{\sqrt{2}}$, where
$|O_{1,2}\rangle$ are eigenstates of $O$ 'internal pointer' observable $Q_O$
with eigenvalues $q^O_{1,2}$. 
For suitable $D$, $O$ Hamiltonian $H_{D,O}$ one can obtain at  $t>t_2$: 
$$
   \Psi_{S,D,O} = \sum a_i|s_i\rangle|D_i\rangle |O_i\rangle
%  a_2|s_2\rangle |O_2\rangle                                   
$$
As easy to see, $D$ states only double $S$ states for this set-up, so 
$D$  can be dropped for the simplicity. In such scheme $S$ directly
interacts with $O$ by means of  Hamiltonian $H_{S,O}$, resulting in the final
state $\rho_{MS}$ and correponding state vector:
 \begin {equation}
   \Psi_{MS} = \sum a_i|s_i\rangle |O_i\rangle
%  a_2|s_2\rangle |O_2\rangle
                                   \label {A72}
\end {equation}
relative to some external RF $O'$.
% From the arguments given above  we can expect that
% $s_i$ eigenstates percepted by $O$ as $Q_O$ eigenvalues $q^O_i$. 
% Considering the general measurement formalism, note first, that
Our aim  is to find find the relation between this state and
 the information acquired by $O$,
 which is quite intricate problem.
In Information Theory,  the  measurement
 of parameters of arbitrary system  $S'$  by an information  system $O^I$
 is the  mapping of $S'$ states set $N_S$
 to  the set $N_O$ of $O^I$ internal states$^6$.
%  In the simple cases   some measured $S'$ parameter
%   $A^S$ becomes correlated with with some $O$ parameter $A^O$
%   via  $S',O$  interaction.
% If  $O^I$ interaction with $S'$ can't be neglected
% If the final $O^I$ and $S'$ states can't be factorized,
 In general  case, which is generic for QM,
the information acquisition by $O^I$ 
can be   described by the formalism of
 systems' self-description$^7$. In its framework,
  $O^I$  considered  as
 the subsystem of larger system $\Xi_T=S', O^I$ with the states set $N_T$.
 This approach gives the most fundamental
and mathematically self-consistent description of the information transfer
 in arbitrary $S'$ measurement called 'measurement from inside'.
 In this case,
the information acquired by $O^I$  about the surrounding objects 
 and $O^I$ itself
 described by $O^I$ internal state  $R_O$ called also
   $\Xi_T$ restricted state or restriction.
% i.e. $O^I$ performs $\Xi_T$ 'measurement from inside'.  
%  In this approach - 'measurement from inside', $N_O$ is the subset
% of $N_T$ - set of $S_T$ states.
 For given $\Xi_T$ system $R_O$ is defined by 
the inference map $M_O$ of $\Xi_T$ state to $N_O$ set, in general,
 $M_O$ should be derived from the first QM principles.
% defines this $O^I$ internal state
The internal $O^I$ state $R_O$ corresponds to $O^I$ internal
 degrees of freedom, which, in principle, can 'record' the incoming
 information.
For example, if $O^I$ is the  atom,  $R_O$ would describe the  state of its
electron shells, which are the functions of the relative coordinates of
 electrons and nucleus.
% which describes the information pattern available to $O$.
% From $N_{T}$ mapping properties
 The important property of inference map $M_O$  
 % $S_T \rightarrow O^I$  inference map
 is formulated by Breuer Theorem: if for two arbitrary $\Xi_T$
states $\Gamma,\Gamma'$
their restricted  states $R, R'$ coincide, then for $O^I$
 this $\Xi_T$
states are indistinguishable, and for any nontrivial $S',O^I$
at least one such pair of states exist $^5$.
%  Under simple assumptions about $S_T,O^I$  at least several such  $S_T$
%  states should exist.
% Contrary to  its  seeming triviality,Breuer theorem stipulates the important
%  information systems features   considered below.
  In classical case, the origin of this  effect   is obvious:
  $O^I$ has less degrees of freedom  than $\Xi_T$ and hence
 can't discriminate all possible $\Xi_T$ states$^7$.
%  because of it some number of $S_T$ indistinguishable states
%  always should exist for any classical $S_T$ and $O^I$ (Svozil, 1993).
   In quantum case,  
  the   entanglement and nonlocality play the additional important role
  in this effect and make its description more complicated.
% The main advantage of Breuer theory is  the proof of
% existence of the complete set of restricted states
% for an arbitrary  $S_T$ structure and correspondingly
% for  any consistent inference map $M_T$.
Despite that $R_O$ are incomplete $\Xi_T$ states,
 they are the real physical states
for $O^I$ observer - 'the states in their own right', as Breuer
characterizes them.
as will be argued, in this approach the information acquired
 by $O$ also    expressed by 
  IP $J$,  main features of IPs described in chap. 2,
 conserved also for $R_O$ states$^8$. 

The mapping relations between
$S'$,$O^I$, $\Xi_T$ states are applicable to MS measurement model
 which  can be also treated as  'measurement from inside' .
 However, Schr$\ddot o$dinger QM dynamics by itself doesn't permit 
to derive
% to derive the restriction ansatz for $S_T$ individual quantum states
% unambiguously and leaves the considerable freedom for the choice of
the inference map $M_O(\Xi_T \to O^I)$ unambiguously, it needs more
detailed development of  formalism described below.
Breuer attempted to avoid this ambiguity phenomenologically, 
 assuming that for  arbitrary  $\Xi_T$
its restricted  state  is equal to the
partial trace of $\Xi_T$ individual state over $S'$, i.e. $R_O$ is
$\Xi_T$ partial state on $O^I$.
% and so coincides with  QM ansatz  for partial $S_T$ states (Breuer,1996).
In our set-up for MS pure state $\Psi_{MS}$  of (3) it gives:
%   $R_O=R^{st}_O$ of (\ref {AA4}) (Breuer,1996). Therefore
\begin {equation}
   R^B_O=Tr_s  {\rho}_{MS}=\sum |a_i|^2|O_i\rangle\langle O_i|
      \label {AA4}
\end {equation}
Plainly, such ansatz excludes beforehand any
kind of stochastic $R_O$ behavior, and this is natural for
Schr$\ddot o$dinger formalism.
% it can be any  consistent fixed map which gives average
% over ensemble reproduces $R^{st}_O$. For the pure case MS
% individual state is  $\Psi_{MS}$ of (2),
%  for  the incoming S  mixture  - Gemenge $W^s$ of (\ref {AA33})
For MS mixed ensemble, induced by incoming $W^s$ ensemble of (1),
the individual  MS state   differs from event to event:
\begin {equation}
\varsigma^{MS}(n)=|O_l\rangle \langle O_l|| s_l\rangle\langle s_l|
  \label {A44}
\end {equation}
where the frequencies of random $l(n)$ appearance in given event $n$
are stipulated by the
 probabilistic distribution $\rm P_l$ $=|a_l|^2$.
% This individual state is  pure and
%  can be initially unknown for $O$ but exists objectively.
% in any event $\varsigma^{MS}(n)$  differs from
%  MS  state (\ref{AA2}).
In this approach
$O$ restricted   state for this mixed ensemble  is also stochastic:
in a given event 
$$
              R_O^{mix}(n)=.\xi^O_1.or.\xi^O_2.
$$
% |O_l\rangle \langle O_l|$   differs   from $R_O$ in any event,
where $\xi^O_i =|O_i\rangle \langle O_i|$
 appears with the corresponding probability $\rm{P}_i$, so that
the ensemble of $O$ states described by the  gemenge
 $W^O_{mix}=\{\xi^O_i, \rm{P}_i\}$.
%       Here $O$ perception
%      is the excitation of some $O$ internal state.
 $R^{mix}_O(n)$ differs formally from $R^B_O$ in any event $n$,
 hence   for the
restricted $O$ individual states the main condition of cited  theorem
is violated.  From that Breuer concluded
that  $O$ can discriminate the individual pure/mixed 
 MS states 'from inside', therefore,
$O$ can discriminate  the individual pure and mixed S states,
it supposedly means that the collapse of pure state doesn't occur$^5$.

However, we find that  the structure of MS observables set (algebra)
together with Schr$\ddot{o}$dinger dynamics
 permit to calculate   MS restriction to $O$ unambiguously,  
 and the obtained results contradict to Breuer conclusion.
 Consider  the measurement of $S$ eigenstate $|s_i\rangle$,
it produces MS individual $\varsigma^{MS}$ states,
 which restriction are $\xi^O_{1,2}$ states with eigenvalues 
 $q^O_{1,2}$.   One can expect that  $O$ identifies  this states  as IP:
$$
        J=J^O_{1,2}=q^O_{1,2}
$$
Really, the
 difference between
 $ \xi^O_i$ states is boolean (classical), because
their mutual relation expressed as: $\langle O_i|O_j\rangle=\delta_{ij}$,
and according to Segal theorem, it corresponds to the relation
 between the classical discrete states
defined on $Q_O$ axe$^4$.
 In addition, in QM formalism $\xi^O_i$ eigenvalues
$q^O_i$ are $O$  real properties, so this hypothesis seems to be well
founded.
The proposed correspondence $\xi^O_i \to J^O_i$
is quite important for our theory, because it establishes
the connection (mapping) between some $O$ quantum states and classical
 IP set $\{J \}$. Generally speaking,
 this set  is the principally different entity from any space of dynamical
states, both quantum and classical,
since its elements $J$ describe the results of $O$ identification 
 of incoming signals$^7$. 

% so the establishment of their
% correspondence with IP $J^O_i$, which describes $O$ perception,
% seems well founded. 

Now we shall use the obtained IP ansatz for the consideration of
 state collapse, in particular, we shall reconsider Breuer conclusions.
 Note  that the formal
difference of two  restricted states doesn't mean automatically
that their difference will be detected by $O$. Such  difference
is the necessary
but not sufficient condition, there should be also some particular
 effect available for $O$ observation,
 which indicate this difference. 
% In our case  it's necessary that  some MS parameters available for
% $O$ observation are different for that states,
% otherwise  this  states would be equivalent for $O$.
%  i.e. the expectation values of all observables available for $O$
% for $O$ (Mittelstaedt,1998).
  For $\xi^O_i$
this are the eigenvalues  $q^O_i$ of observable $Q_O$,
 resulting in  BD of $\xi^O_{1,2}$ for
$O$ and described by IP $J^O_{1,2}$.
% and so $O$ can discriminate $\xi^O_i$ states. 
 Analogously, one should explore  whether some discriminating effect
observed by $O$ can indicate $R_O$ and $\xi^O_i$ difference.
%  the conclusion that $O$ can distinguish $R_O,\xi^O_i$
    The check  of this hypothesis can be performed analogously to
the ansatz described by formulae (\ref {E1}), 
but for the completeness of our proof it's instructive to use the
 alternative operator methods.
 Suppose that $R_O \ne \xi^O_{1,2}$ for $O$
 in BD sense,
% i.e. this is the logical proposition with boolean value $.true.$,
so that their relation should be expressed also by some $O$ parameter
 $g$ with  the values $g_0 \ne g_{1,2}$, correspondingly.
 In QM formalism, such parameter $g$, if it exists, should correspond
 to some $O$ observable $G^O$.
% in particular,  $R_O$
%projector $P^R$ is the possible candidate.
%  Since in our case three $O$ states are involved,
 It should be such $O$ PV observable $G^O$, for which $R_O,\xi^O_i$
are its eigenstates with projectors $P^R$, $P^O_i$ and eigenvalues
 $ g_0,\, g_i$, such that  $g_0 \ne g_i$.
%   But there is no 
% $G^O$ with such properties,  it follows from the fact that 
% each $\xi^O_i$ has one and only projector $P^O_i=\xi^O_i$.
% as any Dirac vector in two-dimensional Hilbert space.
From  Spectral Theorem$^3$ 
an arbitrary Hermitian operator $G^O$, for which
$R_O,\xi^O_i$ are eigenstates,  admits the orthogonal decomposition:
\begin {equation}
      G^O=G'+g_0P^R+g_1 P^O_1+g_2 P^O_2  \label {DD1}
 \end {equation}
% here $g_i$ are an arbitrary real numbers,
here  $G'$ is an arbitrary operator
for which $G'P^O_i=0$, $G'P^R=0$; it should be also $P^RP^O_i=0$.
  But $P^O_i$ constitute
the complete algebra of projectors in $\cal H_O$ corresponding to 
the  orthogonal unit decomposition: $\sum P^O_i=I$,
 and so:
\begin {equation}
     P^R= P^RI=P^R \sum P^O_i=\sum P^R P^O_i=0,  \label {DD3}
 \end {equation}
% it excludes for $R_O$ the possibility to be the  eigenstate of any $G^O$.
% Consequently it's impossible for $O$ to discriminate $R_O$
% from $\xi^O_i$ in the individual events.
Hence $R_O$ can't possess the independent projector in $\cal {H}_O$  and such
nontrivial $G^O$
 doesn't exist. From the correspondence between the states and their projectors
follows that  $R^B_O$ of (5) isn't proper ansatz for
$\Psi_{MS}$ restriction $R_O$, the only solution is to accept that
in any event $P^R=P^O_{i}$ and correspondingly:   
 \begin {equation}
         R_O=.\xi^O_1.or.\xi^O_2.   \label {B3}
\end {equation}
i.e. it coincides with $R_O^{mix}$ as the individual state.

For pure MS ensemble the  expectation value $\bar{Q}^l_O$ for any natural $l$
  can be calculated without the use of Reduction Postulate
 from Graham-Hartle theorem$^{11}$,
based on quite loose assumptions.
 To reproduce this $\bar{Q}^l_O$ values,
 $O$ should observe the  collapse of pure MS state
to one of  $q^O_i$ at random with probability $\rm{P}_i$ $=|a_i|^2$, i.e.
the ensemble of $O$ states described by the  gemenge
 $W^O=\{\xi^O_i, \rm{P}_i\}$. It induces the corresponding $O$ IP ensemble 
$Z^O=\{ J^O_i, \rm {P}_i \}$.
Eventually, the inference map $M_O$ for
$\Psi_{MS}\rightarrow R_O$ restriction is stochastic,
% and results in  the  subjective state collapse  observed by $O$.
 we don't present here $M_O$ ansatz  in the analytical form, which
can be easily derived from the previous calculations but is rather tedious.
%
% the alternative derivation of MS restrictions to $O$ will be described
Our studies show that POV generalization of standard QM PV observables
don't change our conclusions. The reason of it is that POV parameters
respond to the nonorthogonal unit decomposition, yet BD can exist
only between mutually orthogonal states, hence it can be shown that
 there is no $O$  POV  observable
which responds to BD between $R_{O}$ and $\xi^O_i$.  
 In the regarded  case, only the parameters corresponding
 to nonlinear operators can 
 establish BD between this states.

 Thus, our assumption about the role of information constraints
 in state collapse is proved
 formally, here we shall discuss in more qualitative
terms  the physical mechanism of such stochasticity in the measurements,
 because it  presents the significant interest for the study of
QM foundations.   
% which doesn't exploit any phenomenological assumptions.
 Remind  that in QM formalism two kinds of uncertainties 
exist:  suppose that for some state $\rho$ the value $\tilde g$  of 
 observable $G$ lays in the interval $g_{min} \le \tilde g \le g_{max}$,
 Then, depending
on  $\rho$, it can be either
the stochastic value, i.e. objectively $\tilde g=g_i$ 
 with some probability $P'_i$, if  $\rho$ is a mixed state,
 or it can be truly uncertain (fuzzy) value  $\tilde g$ for pure $\rho$.
The difference between this two states revealed by 
   'interference term' (IT) observables, 
 which demonstrate the presence of
$\tilde g $ superposition.
%  As we argue below, the (im)possibilityto differ this
% two kinds of QM uncertainty is the essence of Measurement Problem.  
For MS entangled states no $O$ observables are sensitive to it,
it can be only joint $S$,$O$ observables $B^{MS}$. As the example,
consider the symmetric IT for MS:
\begin {equation}
   B=|O_1\rangle \langle O_2||s_1\rangle \langle s_2|+j.c.
    \label {AA5}
\end {equation}
% which characterizes $O$,S nonclassical correlations - i.e.
% their entanglement.
Being measured by external RF $O'$ via its interaction with $S$, $O$,
 it gives $\bar{B}=0$ for any  $|s_i\rangle$ incoming mixture,
 but  $\bar{B}\neq 0$  for  entangled MS states of (4).
For example, for the  incoming symmetric $S$ state $\psi^s_s$
 with $a_{1,2}=\frac{1}{\sqrt{2}}$, the corresponding
$\Psi^s_{MS}$ is $B$ eigenstate with eigenvalue $b_1=1$.
% Qualitatively  for  the initial state  $\psi^s_s$,
% as the result of S,$O$ interaction,
% $S_z$ observable  mapped to $Q_O$, whereas
% $S_x$ to $B$  observable.
% On the whole IT observables are nonlocal
% and  constitutes the large class  $\{B^{MS}\}$ which will be used below.
 However,  $B$ value  can't be directly measured by $O$
 'from inside', at least simultaneously with $S_z$, because they
 don't commute$^5$.
%   such distinction between pure and mixed  $O$,S state  is
%  unavailable for $O$ directly.
 In addition, when $S,O$ interaction finishes, $S$ can  become free
 particle again, and so  the joint $S,O$  observables can become
 unavailable for $O$ in a short time.
% but only its own internal observables, as the restriction of MS
% to $O$ dictates.
% the nocommutativity: $[B,S_z] \neq 0$, 
% principal possibility for $O'$ to  measure $B$
% and send the information to $O$ doesn't change
% the situation, because such  measurement by $O'$ is incompatible
% in general with $S_z$  measurement by $O$.
 From the same reasons  the whole set of IT observables 
  $\{B^{MS}\}$ is unavailable for $O$ during $S_z$ measurement,
but only some $O$ internal observables.
Note that, in general, the pure/mixed MS states with the same
$\bar{Q}_O$ can be discriminated
 only statistically, since their   distributions of $B$ values (or
other $B^{MS}$)
overlap.  Namely, for $\Psi^s_{MS}$ the probability  $P_B(b_{1,2})=.5$
for such mixture, so its $b$ distribution intersects largely
 with $b$ distribution for $B$ eigenstate $\Psi^s_{MS}$. 
 Consequently, even $O'$  can't 
% even if to imagine that $B$ will be available for $O$,
% it wouldn't permit to
discriminate  the pure/mixed MS states in a single event,
 but only statistically for MS ensemble with $N \to \infty$.
Overall, it follows  from this analysis that no IT
observables are available for $O$, and so one can expect
 that $O$ can't discriminate
the pure and mixed $O$ states, for which $q^O$ is 'smeared' inside
the same uncertainty interval.
 Yet we know that $R^O_{mix}$ ensemble
$W^O_{mix}$ can reproduce the same $\bar{Q}_O^l$ values as  $\Psi_{MS}$
ensemble.
 Because of this reasons, $R_O$ ensemble $W^O$
 should coincide with the mixed ensemble $W^O_{mix}$.
 Consequently, for $\Psi_{MS}$ the genuine
 $q_0$ uncertainty can be  released in $O$ RF  
only in the form of $q_0$ randomness inside its uncertainty interval
$\{q^O_1,q^O_2\}$.
Roughly speaking, $\Psi_{MS}$   induces
 $O$ signal, which in each event put some income into the resulting
  $\bar{Q}_O^l$ value, so that
 their ensemble should produce
the  proper $\bar{Q}_O^l$ values.
But it turns out that such signal can  be only stochastic,
no ensemble of  identical signals with demanded  properties exist.
It seems that our results constitute the kind of no-go theorem
 for the observation of $|s_i\rangle$ superpositions by $O$.

It's well known that the decoherence of pure states by its environment $E$
is the important effect in quantum measurements$^{1,2,10}$,
 we find yet that $O$ decoherence by  $E$
doesn't play the principal role in our theory. However, 
its account stabilizes the described collapse mechanism additionally and  
defines unambiguously the preferred basis {PB} of $O$ stable 
final states $\{\xi^O_i\}$ exploited here.
Really, for the specially chosen Hamiltonian of $O$,$E$ interaction$^{10}$,
 one obtains that MS,$E$ final state is:
$$
    \Psi_{MS,E}=\sum a|s_i\rangle|O_i\rangle \prod\limits^{N_E}_{j}
           |E^j_i\rangle 
$$  
where $E^j$ are $E$ elements,
 $N_E$ is $E^j$ total number. As easy to demonstrate, if an arbitrary
$O$ pure state $\Psi_O$ is produced, it will  also qecohere in a very
 short time into the analogous $|O_i\rangle$
combinations, entangled with $E$,
 so that $O$ can practically percept
only   $|O_i\rangle$ final states. Such $O$ PB can be derived
 from other more subtle
arguments, yet $O$ decoherence even for quite small $N_E \sim 2 \div 4$
already defines it effectively.

We conclude that standard Schr$\rm\ddot{o}$dinger QM formalism 
 together with the theory of quantum systems' self-description
permit to obtain the collapse of the measured pure state
without implementation  of independent Reduction Postulate
into QM axiomatics.   
 As was shown, in this approach the main
source of stochasticity  is  the principal
constraint on the transfer of  specific information in S$\,\to O$
 information channel.
  This  information, unavailable for $O$,
characterizes the purity of S state$^8$,
 because of it, $O$ can't discriminate the pure/mixed S states.
 As the result of this
information incompleteness,
the stochasticity of measurement outcomes appear,
 which is the analog of fundamental
 'white noise'..
 In addition, the formalism of systems'
self-description  permits to resolve also the old problem of
Heisenberg cut in quantum measurements, by the inclusion of the
information system into quantum formalism properly and on equal terms
with other MS elements$^9$. Of course, the most exciting and controversial 
question is whether this theory is applicable to the observations made by human
observer $O$, in particular, whether IP $J$ describes the true $O$ 
'impressions' about their outcomes ? This is open problem,
but since our theory is based on standard QM premises, and at the
microscopic level the human brain should obey QM laws, we believe
that the answer can be positive. Note that in our theory the brain
or any other processor plays only the passive role of signal receiver,
the real effect of information loss, essential for collapse,
occurs 'on the way', when the quantum signal passes
 through the information channel. 
% It agrees qualitatively
% with Heisenberg  uncertainty relations: $O$ measures $S_z$, but
% S state purity characterized by simple function of $\bar{S}_x,\bar{S}_y$,
% and this S observables don't commute.
The  interesting feature of this theory
 is that the same MS  state can be stochastic
in $O$ RF, but evolve linearly in $O'$ RF.
In particular, $\Psi_{MS}$ restriction to $O$ in $O$ RF is  stochastic 
state $R_O$ of (9), yet in $O'$ RF $O$ partial state is $R^B_O$ of (4),
i.e. is  the 'weak superposition'.
 The detailed explanation
of this effect is given by the unitarily nonequivalent representations
admitted in  Algebraic QM$^8$. Here we notice only
that $O$ and $O'$ deal with different sets of MS observables, and so  
the transformation of MS  state between them can be nonunitary.
 Obtained results agree well
with our calculation in $C^*$ Algebras formalism$^8$, in that approach
the inference map $M_O$ is the operator restriction of MS observable algebra
to $O$ (sub)algebra.
\vspace* {-6pt}
%\section* { References}

\end {document}